%% file: ms.tex
\documentclass[sigconf]{acmart}
\usepackage{booktabs} 
\usepackage{subcaption}
\usepackage[ruled]{algorithm2e} 

\setcopyright{acmcopyright}

\acmDOI{0000001.0000001}

\begin{document}
\title{Influential User Subscription on Time-Decaying Social Streams}

\author{Xin Yang}
\affiliation{%
  \institution{School of Information, Renmin University of China}
  \state{Beijing}
  \country{China}
}
\email{yangxin09@ruc.edu.cn}
\author{Ju Fan}
\affiliation{%
  \institution{School of Information, Renmin University of China}
  \state{Beijing}
  \country{China}
}
\email{fanj@ruc.edu.cn}

\input{notations}
\input{abstract}

%
%


%
%


\maketitle

\input{intro}
\input{problem}

\input{timedecay}

\input{prefixtree}

\input{experiment}

\end{document}

%% file: notations.tex
\newcommand\argmax{\operatorname{arg\,min}}

\newcommand{\est}{\ensuremath{e}\xspace}
\newcommand{\estset}{\ensuremath{S}\xspace}
\newcommand{\N}{\ensuremath{N}\xspace} 
\newcommand{\action}{\ensuremath{a}\xspace}
\newcommand{\query}{\ensuremath{q}\xspace}
\newcommand{\tpath}{\ensuremath{P}\xspace}
\newcommand{\ests}{\ensuremath{E}\xspace}
\newcommand{\sn}{\ensuremath{SN}\xspace}
\newcommand{\V}{\ensuremath{\mathcal{V}}\xspace}
\newcommand{\E}{\ensuremath{\mathcal{E}}\xspace}
\newcommand{\edge}{\ensuremath{e}\xspace}
\newcommand{\prf}{\ensuremath{P}\xspace}
\newcommand{\topic}{\ensuremath{T}\xspace}
\newcommand{\Q}{\ensuremath{Q}\xspace}
\newcommand{\A}[3]{\ensuremath{A_{#1 \rightarrow #2}^{#3}}\xspace}
\newcommand{\stream}{\ensuremath{S}\xspace}
\newcommand{\infset}[1]{\ensuremath{\mathcal{I}(#1)}\xspace}
\newcommand{\weight}[1]{\ensuremath{\mathcal{w}(#1)}\xspace}

\newcommand{\syminfl}{\ensuremath{f}\xspace}
\newcommand{\infl}[1]{\ensuremath{f(#1)}\xspace}
\newcommand{\margin}[2]{\ensuremath{\Delta(#1|#2)}\xspace}

\newcommand{\psim}{\textsc{PSIM}\xspace} 
\newcommand{\algoPSIM}{\textsc{PSIM}\xspace}
\newcommand{\algoTimeDecay}{\textsc{TimeDecay}\xspace}
\newcommand{\algoSieveStreaming}{\textsc{SieveStreaming}\xspace}
\newcommand{\algoPrefixSieve}{\textsc{PrefixSieve}\xspace}
\newcommand{\algoModify}{\textsc{Modify}\xspace}
\newcommand{\algoFindPath}{\textsc{FindPath}\xspace}
\newcommand{\algoClear}{\textsc{Clear}\xspace}

\newcommand{\IC}{\textsc{IC}\xspace}
\newcommand{\SIC}{\textsc{SIC}\xspace}

%% file: abstract.tex
\begin{abstract}
  Influence maximization which asks for $k$-size seed set from a social network
  such that maximizing the influence over all other users (called inlfuence spread)
  has widely attracted attention due to its significant applications in viral
  markeing and rumor control. In real world scenarios, people are interested in
  the most influential users in particular topics, and want to subscribe the
  topics-of-interests over social networks. In this paper, we formulate the
  problem of influential users subscription on time-decaying social stream, which
  asks for maintaining the $k$-size inlfuential users sets for each topic-aware
  subscipriton queries. We first analyize the widely adopted sliding window model
  and propose a newly time-decaying influence model to overcome the shortages when
  calculating the influence over social stream. Developed from sieve based streaming algorithm,
  we propose an effecient algorithm to support the calculation of time-decaying influence
  over dynamically updating social networks. Using information among subscriptions,
  we then construct the Prefix Tree
  Structure to allow us minimizing the times of calculating influece of each update
  and easily maintained. Pruning techniques are also applied to the Prefix Tree to
  optimize the performance of social stream update. Our approch ensures a
  $\frac{1}{2}-\epsilon$ approximation ratio. Experimental results show that
  our approach significantly outperforms the baseline approaches in effeciency and
  result quality.
\end{abstract}

%% file: intro.tex
\section{Intorduction} \label{sec:intro}

Influence maximization, which asks for k-size set of users in a social network
maximizing the influence speard over all users. Online social networks, like Facebook
and Weibo, have boosted researches on the influence maximization problem due to
its potential commercial value, such as viral marketing \cite{?}, rumor control, and
information monitoring \cite{?}.


In real world social networks, users have topics or keywords indicating their fields
of interests, e.g., hashtags of Twitter, subreddits of reddit, etc.. A user related to
certain keywords or topics will more possibly influence and be influenced by other
users. For example, a user who are interested in basketball will participate in
discussion of subreddits such as \texttt{sport}, \texttt{basketball} and
\texttt{MBA}, the related keywords of this user can be represented as
$\{\mbox{\texttt{sport}}, \mbox{\texttt{basketball}}, \mbox{\texttt{MBA}}\}$.
In this case, recording numbers of subscription queries,
people can subscribe the most influential users in particular areas of
interests. For example, one can subscribe $\query = \{ \mbox{\texttt{nerual learning}},
\mbox{\texttt{machine learning}}\}$ to keep track of the users who are most influential
in the area of machine learning over time. In this paper, We formulate problem of influential
user subscription on time-decaying social stream (or \psim problem for short), which
asks for $k$-size seed set of every subscription queries on dynamic social stream.


The topic-aware influence maximization can be applied to many real world scenarios.
For example, advertiser who has limited budget hope that its advertisement will be
push to the most influential users on some particular topics.
The advertiser will hope the users who recieve the information will be interested
in some particular topics, who will most likely repost the advertisement.
Therefore, the social network companies will want the \psim subscriptions for
certain topics or keywords, and locate these users for the advertisers.


Our formulation is different from the traditional online influence maximization
in three ways. First, as the influence possibility between two users on social networks decays
over time, i.e., a more reccent response action (e.g., \emph{repost}, \emph{comment} and
\emph{cite}) between a pair of users infers stronger influence between the two
user than a more previous action, we propose \emph{time-decaying} influence
model in this paper to meet this intuition. Second, we take into consideration
keywords or topics of users. The most influential users are constrained by both
influence and related keywords. Third, we aim at solving the \psim problem of
hundreds of thousands of subscription queries, which means our algorithm should
be effecient enough to support the online queries.
However, the \psim problem is
NP-hard. We develop a approximating algorithm from sieve based streaming algorithm to meet
the requirement of both online subscription pushing and high quality.

%


To achieve the effeciency and quality requirement for answer the \psim problem, we
first develop the naive sieve based streaming algorithm to support the fast calculation
of time-decaying influence model over envolving social networks. In order to improve
the performace, we propose a Prefix Tree Structure which meets the the following
properies:
(1) every candidate sets are stored only once, thus the marginal influence of same
candidate sets will not be calculated repeatedly,
(2) downward closure property can be applied to the Prefix Tree to minimizing the
times of calculating marginal influence, and
(3) can be easily updated as the actions of social stream arrives in sequence.
We propose an efficient streaming update algorithm based on the Prefix Tree Structure and
design three pruning conditions which are applicable on the Prefix Tree to
avoid unnecessary marginal influence calculation of each update.
Our approch ensures a $\frac{1}{2} - \epsilon$ approximation for \psim problem,
where $\epsilon$ is the approximation ratio subject to $\epsilon \in (0, 0.5)$.

To summarize, we make the following contributions.

\begin{itemize}
  \item We propose a time-decaying influence model, which outperforms sliding window
  models in both quality and efficiency.
  \item We propose Prefix Tree Structure which can be easily updated for minimizing
  the times of calculation of marginal influence of each update. We also propose
  the pruning techniques over Prefix Tree to avoid unnecessary influence calculation.
  We develop algorithm based on the Prefix Tree Structure which returns the results
  for \psim problem when an approximation ratio $\frac{1}{2} - \epsilon$.
  \item Experimental results on real datasets show our algorithm significantly
  outperforms state-of-the-art approaches.
\end{itemize}

The rest of this paper is organized as follows. Section \ref{sec:problem} formulate
the \psim problem. We then discuss the time-decaying model and develop the
sieve based streaming algorithm to support this influence model, which will be
discussed in Section \ref{sec:timedecay}.
In the following seciton, i.e., Section \ref{sec:prefixtree}, we propose the Prefix
Tree Structure and discuss the streaming update over the Prefix Tree. The pruning
techniques will be also discussed in this section. Section \ref{sec:experiment}
reports the experiment results. We conclude our work and result in \ref{sec:conclusion}.


%% file: problem.tex
\section{Problem Definition} \label{sec:problem}


\subsection{Time-Decaying Social Streams}

A \emph{Social Network} is a graph reveals some kind of influence relation
between the users on an online society, e.g., Twitter, DBLP, etc.. A social graph
can be formulated as $\langle \V, \E \rangle$,
denoted as $\sn$, where $\V$ is the vertex set and $\E$ is the edge set. For each
edge $\edge \in \E$, we define the tail end user $u_r$ as influencer, and the head
end user $u_e$ as influencee.

\emph{Social Stream} is the sequence of \emph{actions}
generated by the users from a beginning time to the current time $t_0$, which is
denoted by $\stream = \{a_1, a_2, \ldots, a_m\}$.
In particular, each $a_{i} = \langle u_e, t_e, u_r, t_r \rangle$ represents that a user
$u_e$ performs an response at time point $t_e$ to another user $u_r$'s social
activity. In different senarios, the response refers to different kind of social
actions, e.g., a \emph{citation} of a former research paper in
bibliography networks, or \emph{post} on Twitter. The action represents one time influence
from $u_r$ to $u_e$, therefore, the action performer $u_e$ and $u_r$ are
called \emph{influencee} and \emph{inlfuencer}, respectively. We consider there is
a \emph{influence relation} from the influencer and the influencee.

Next, we provide some examples to illustrate the intuitions of the above-defined
action and influence.

\begin{itemize}
	\item In an online social network, such as Twitter and Reddit, users either
  create posts (e.g., tweets) or respond to others' posts (e.g., retweet, comment,
  etc.). In such a scenario, the influence relation can be extracted from the
  poster $u_r$ and the respondent $u_e$. Each post can be formalized as an action $a$ with
  influencee $u_e$, influencer $u_r$, the response time $t_e$ and the responded
  performance time $t_r$.
	\item A bibliography database, such as DBLP~\cite{xxx}, maintains citations
  among research papers. In such a database, each paper can be considered as a set of
  action $\{ a_i\}$ consisting of the influence relation from the authors of the citing
  paper to the authors of the cided paper. In this scenario, the influencer time and
  influencee time are the publish time of the research papers.
\end{itemize}

From the aforementioned examples, we can see that social action influence relations
 are general formalizations that capture many real-world scenarios.

Each action is given a \emph{time-decaying weight} based on the influencer time and
influencee time of the action, denoted as $\weight{a}$.
The intuition of the weight is that recent actions are more significant than old
ones to capture the strength of the influence between two vertexes.
The time-decaying weight will be further discussed in the following part.


\subsection{User Influence Model}


As mentioned in previous part, social stream is time-sensitive. The strength between
the influencer and influencee revealed from an influence relation is depend on the
influencer time $t_r$ and influencee time $t_e$: decreases by the time elapse from
$t_r$ and $t_e$ to current time $t_0$.

Sliding window models are the most widely-adopted model to handle the time
sensitivity of streaming data. There are two kinds of sliding window models:
sequence-based and time-based~\cite{slidingwindow}. The sequence-based mothod
maintains a size-$N$ sliding window over the lasted items in the social stream;
while the time-base mothod maintains a flexible length window with fix-length of time.
For the sequence-based time sliding window, the length of time will be changing
when the window slides, however, in real-world streaming scenarios, the number of
actions per timestamp can be largely diversed. Therefore, this method cannot control
the length of time window flexibly. Time-based sliding window can maintain a fixed
length of time window of interest, however, it regards the actions in the time window
with same weight. Therefore, it cannot explicitly reveals the importance of the
actions within the sliding window. A common shortage of both of these sliding window
models is that they throws away some information of the previous actions, which
can leads to bias of the of the calculation of influence of the social network.

In this paper, we assign each action a time-decaying weight, which is non-decrease
with the influencer time and the influencee time of the action. In this paper, we
focus on the the commonly used \emph{exponential decay function}, i.e.,
\begin{equation}\label{eq:a-weight}
	\weight{a} = e^{-\lambda \cdot [(t_e - t_0) + (t_r - t_0)]},
\end{equation}
\noindent where $t_0$ is the current time, and $\lambda > 0$ is a parameter for
controlling the decay speed (aka. decay constant).
Here we can see that the weight of a action increases as the influencer time $t_r$ or
influencee time $t_e$ is getting closer to current time. It is coresponded to the
sensitivity of social streams.

There can be multiple influence relations between a pair of users $u_r$ and $u_e$, the
collection of these actions is denoted as $A_{u_r \rightarrow u_e}$. For each
action $a \in A_{u_r \rightarrow u_e}$, there is a coresponded weight $\weight{a}$.
We consider the influence between $u_r$ and $u_e$ as the largest weght among these
coresponded weights, i.e.,
\begin{equation}\label{eq:inf-u-to-v}
  \infl{u_r \rightarrow r_e} = \max_{a \in A_{u_r \rightarrow u_e}} \weight{a}.
\end{equation}
\noindent The intuition is to maximize the influence ``coverage'' over each node.
In weighted \emph{k set coverage} problem \cite{setcoverage}, which asks for k elements in a
collection of sets which maximize the total coverage, the coverage over each vertex
will be determined by the maximun weight of all the sets covering it, and the total
coverage of a collection of set is determined by summing up all the influence over
each vertex in the universe set. In our case, the influence of
each node is considered as a ``set'' over the set of nodes in the universe set $\N$,
where the weight
from an action $\action=\langle u_e, t_e, u_r, t_r\rangle$ provides a partial
coverage to the influence of $u_r$ over $u_e$. Therefore, the total coverage of
the influence of user $u_r$ over $u_e$ will be the equal to the maximum coverage
among the actions from $A_{u_r \rightarrow u_e}$.

With the influence of between to users, we can further discuss how we determine
the influence of a set of users $S$ to a user $v$.
Similarly, in set coverage point of view, the influence of $S$ is equivilent to
to the maximan influence of each $u \in S$. Therefore, the influence of the $S$ to user
$v$ can be determined by the maximum influence of the influence of each user $u \in S$.
Therefore, the influence of a set of users $S$ to $v$ is defined as
\begin{equation}\label{eq:inf-S-to-v}
  \infl{S \rightarrow v} = \max_{u \in S} \infl{u \rightarrow v}.
\end{equation}

By the abovementioned definitions of influence to user, we can determine the total
influence of a certain user or a set. We define the set of users inlfuenced by user
$u$ as \emph{influence set} of \emph{target} $u$, denoted as $\infset{u}$. Similarly,
the influence set of a target collection of users $S$ can be denoed as $\infset{S}$. For each user
$v$ influence sets, the influence from target user or set of users to $v$ is non-zero, while the influence
from $u$ to other users is zero. So we can determine the total influence of target
by summing up all the influence of target over each nodes in the influence set, i.e.,
\begin{align}
  &\infl{u} = \sum_{v \in \infset{u}} \infl{u \rightarrow v} \label{eq:total-inf-u} \\
  &\infl{S} = \sum_{v \in \infset{S}} \infl{S \rightarrow v} \label{eq:total-inf-S}
\end{align}

\begin{figure}[t!]
	\includegraphics[width=\linewidth]{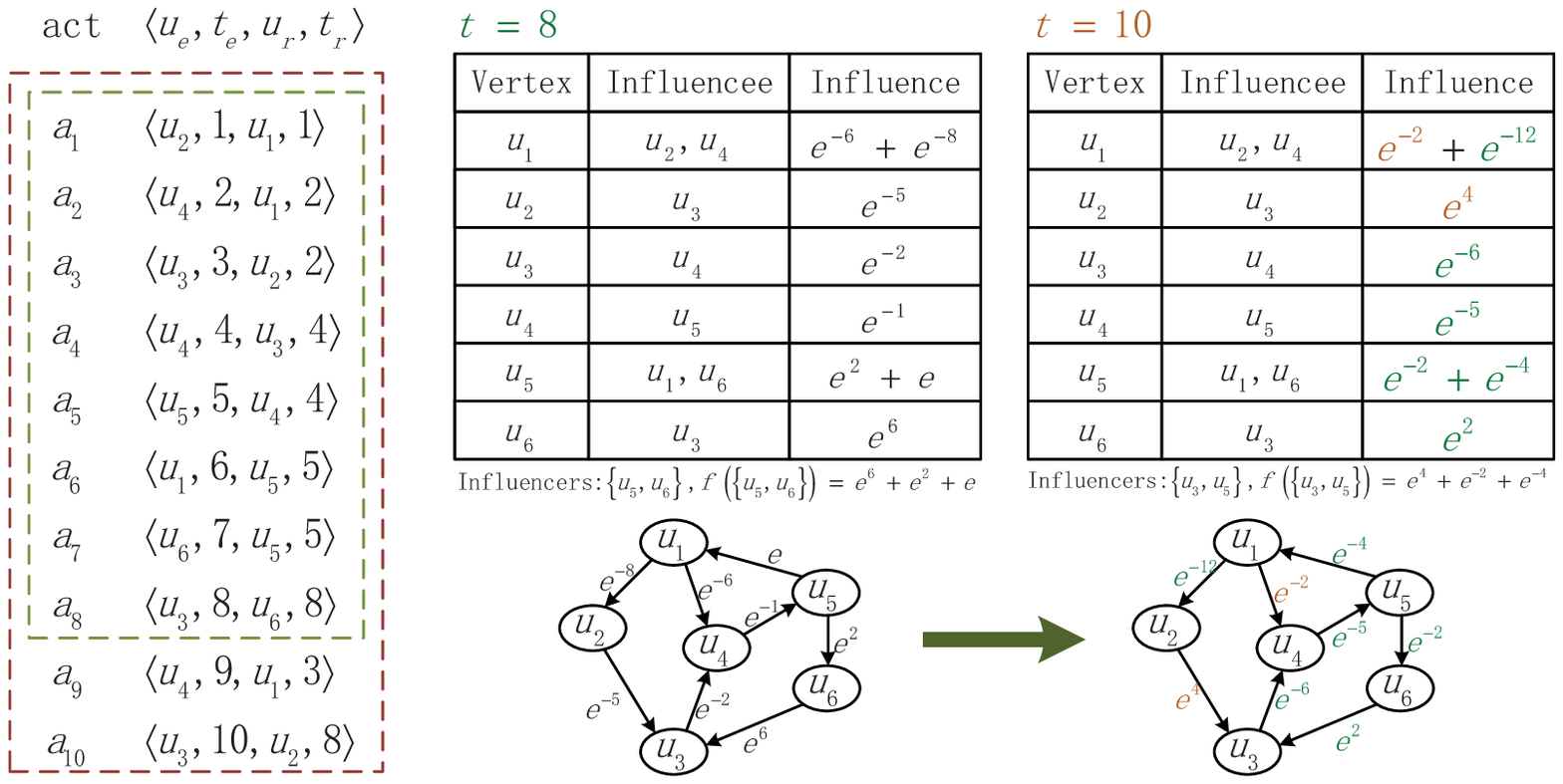}
	\caption{Time decaying model effect over time}
	\label{fig:timedecay}
\end{figure}

The effect of the time-decaying model over time is shown in Fig \ref{fig:timedecay}

\subsection{Influential User Subscription}

Our work focuses on topic-aware Inlfuence Maximization subscription queries.
Users are allow to subscription certain keywords of interests, and the \psim
problem is to find out the most influenctial within these keywords. In this part,
we will first discuss the user profile, with determine the keywords related to
users. We then define the influence maximization subscriptions. Finally, we will
formulate the \psim problem.

\noindent \textbf{User Profile. } In real world social networks, users will have
topic keywords suggesting its topic-of-interests. These keywords, e.g. tags, can
extracted from either the post's content or catagory. For example, A reddit user
whose is interested in basketball will participate in numbers of discussion in
\texttt{basketball}, \texttt{sport}, \texttt{MBA} subreddits, we can therefore
extract the user's set of keywords from the subreddits he most frequently participates
in, i.e. $\{\mbox{\texttt{basketball}},\mbox{\texttt{sport}},\mbox{\texttt{MBA}}\}$.
This set of keywords are called the \emph{profile} of a user, which can be denoted
as $\prf_u = \{ \topic_1, \topic_2, \cdots, \topic_{p_u}\}$, where $\topic_i$ are
the extracted keyowrds of user $u$.

\noindent \textbf{Influential User Subscription. } A subscipriton can be several
keywords-of-interests of which people want to know the most influencers. A subscription
query can be denoted as $\query = \{ \topic_1, \topic_2, \cdots, \topic_{Z_{\query}}\}$.
A query will be corelated to numbers users in the the social network. We determine
the corelationship of subscipriton query $\query$ and user $u$ according to the keywords
sets subsumption, i.e. a user $u$ is related to query $\query$ if $\query \subset \prf_u$.
The intuition is that \ldots

Now, we are ready to define our problem of influential user subscription on
time-decaying social stream (or the \psim problem for short) as follows.

\begin{definition}
  (\textsc{The \psim Problem}) Given a set of subscription queries
  $Q = \{ \query_1, \query_2, \cdots, \query_m\}$ and s social stream $\stream$,
  the problem maintains a size-k user set having the maximum influence with respect
  to each subscription query $\query \in Q$, i.e., $R_q = \arg_{S:|S| < k}\max \infl{S}$
  as stream $S$ continuous updates in time order.
\end{definition}

%% file: timedecay.tex
\section{Time-Decaying Influence Maximization}

\subsection{Time-Decaying \algoSieveStreaming} \label{sec:timedecay}



As far as we know, state-of-the-art influence maximization algorithms can't support
time-decay model, instead, most of them are based on sliding window models. In this
part, we focus on developing a sieve based algorithm \algoSieveStreaming to support
the time-decaying influence model.

\algoSieveStreaming is a set streaming algorithm which finds
the most coverage set by just scanning through set stream only one time.
A $(\frac{1}{2} - \epsilon)$-approximation is ensured by this method.
The former application of \algoSieveStreaming in Influence Maximization is
based on sliding window model \cite{SIM}. In this section, we will improve it to
support time-decaying influence model.

\subsection{Niave Time-Decaying \algoSieveStreaming}

\input{sievestreaming}

The psudocode for \algoSieveStreaming is shown in Algorithm~\ref{algo:sievestreaming}.
The basic idea is to generate a series of estimations $E = \{ e:e = (1 + \epsilon)^i,
s.t.~m \leq e \leq 2km, i \in \mathbb{Z} \}$, where $\epsilon$ is the approximation
ratio, $k$ is the maximun seed set size, and $m$ is the largest set so far in the
process of scanning the set stream. The optimal result of the set streaming
problem lays in the arrange of $(m, 2km)$, for the optimal result will not smaller
than $m$, which is the influence of a single set, while it can not exceed $2km$,
which can only happen when there are $k$ disjoint set with influence $m$.
Each of the the estimation is corelated to a candidate collection of sets, denoted as
$S$, which is originally emptysets.
When a new set is being scanning in the set stream, the algorithm will calculate the
marginal coverage of the current set $s$ w.r.t. each candidate collection $S$,
and compare the value to the sieve conditon, which is

\begin{equation}
  \margin{s}{S} \geq \frac{\frac{e}{2} - \infl{S}}{k - |S|}
\end{equation}

where $\margin{s}{S}$ is the marginal coverage of $s$ w.r.t. $S$, and $\infl{S}$
is the coverage of candidate collection $S$. In the next step, the new-coming
set $s$ will be inserted to the candidate collections which satisfies the sieve
condition, while nothing will be done for those do not satisfy the condition.
The algorithm ends by selecting the collection with maximun coverage among the
candidates and return it as the answer to the most coverage problem.

At each iteration, \algoSieveStreaming will update the maximal set and adjust the
range of its estimations.

The \algoSieveStreaming can be trainslated when applied to Online Influence
Maximization Problem, presented by \cite{SIM}.
In this trainslation, the coverage of a set is replaced by a influence of a user
$\infl{u}$, the marginal coverage $\margin{s}{S}$ is converted to marginal influence
$\margin{u}{S}$, the candidate collections of sets therefore be converted to candidate
user set.

To improve \algoSieveStreaming to allow it to support time-decaying influence model,
a straight-forward idea is to record the future estimations in advance, i.e., $(1 + \epsilon)^i \cdot
e^{2 \lambda \Delta t}$, where $\Delta t$ is the difference between the future time and current time.
In this way, we can calculate the future candidate sets for future estimations.
And When the base time $t_0$ changes, the current estimations will be expired and erased out, the rest
of the estimations will decay by the decaying factor, which leads the estimations of new time to be
exactly the estimations of current time.
In this way, we can modify \algoSieveStreaming to support the time-decaying influence model.

But too many estimations will slow down the streaming process, we therefore should determine which is
the minimum number of future estimations we should store in advance.
By \algoSieveStreaming, we know that the optimal result to a subscription $\query$ should be in the
range of $(m, 2km)$, where $m = \max_{u \in \query} \infl{u}$.
The estimations smaller than $m$ is not necessarily stored, as we know that there is at least one user
will influence of $m$, while the estimations larger than $2km$ will all be empty sets, for there's no
such user the influence of which is larger than $\frac{2km}{2k} = m$.
Therefore, the estimations-in-advance should be in the range of $(m, 2km)$, i.e.
$m < (1 + \epsilon)^i \cdot e^{2 \lambda \Delta t} < 2km$.
As the influences decay by time, the estimations of far future will decay dramatically and hard to detect.
Therefore, we have a minimum detecting threshold $\tau_d$, a set with influence smaller that $\tau_d$ can
be regarded as zero. Therefore, the future estimations should subject to $\tau_d \cdot e^{2 \lambda \Delta t}
< (1 + \epsilon)^i \cdot e^{2 \lambda \Delta t} < 2km$. Therefore, $\Delta t$ is subject to
$\tau_d \cdot e^{2 \lambda \Delta t} < 2km$, and equally,

\begin{equation}
  \Delta t < \frac{1}{2 \lambda} \log{\frac{2km}{\tau_d}},
\end{equation}

which is the upper bound of $\Delta t$.

When the timestamp changes, i.e., $t_{cur} > t_0$, we can erase the timestamps of
current time and their related candidates, and decay the rest estimations and the
influence of their corelated candidate sets by decay factor $e^{-2 \lambda (t_{cur}-t_0)}$.
The resulting estimations are those of new timestamp.

\subsection{Estimation Shift and Lazy Time Decaying}

The naive implementation of time-dacaying \algoSieveStreaming will be time-consuming
for it generates many times more estimations than the orignal algorithm.


Therefore, we hope to make use of the estimations of a former timestamp to generate
the estimations of current time. The basic idea of this improvement is to convert
the estimations of previous timestamp, and make sure they cover the entire range of
possible optimal result for the \psim problem. In the following part, we proved a
theorem which can help us easily generate estimations from the those of previous
timestamp.

On the other hand, in frequent update social networks,
in which the current timestamp changes much more faster than other social networks,
the time-decay will be called frequently and therefore be time-consuming.
However, we can't use exponential increased influence to replace the exponential
decay influence because of the limited data range.
We therefore propose \emph{lazy time decaying strategy} to cut down the overhead costs.
The basic idea is to set a threshold based on data range of differnet systems. The
time decay process will be call only when the maximum influence exceeds the pre-set
threshold. It can be shown in the section~\ref{sec:experiment} that the lazy
time decaying strategy exceeds the naive time decaying process for two orders
of magnitude.


\subsubsection{Estimation Shift}

Instead of keeping track of eveary estimations of optimal result of $(1+\epsilon)^i$,
we record of estimations with value of $b (1+\epsilon)^i$, where b is a random
base parameter smaller than $\max_{u \in \query} \infl{u}$. We prove the following
theorem to guarantee the correctness of our change.

\begin{theorem}
  Given subscription $\query$, for any $b < \max_{u \in \query} \infl{u}$,
  if we denote $\max_{u \in \query} \infl{u}$ as $m$ and generate the set of estimations
  as $\ests = \{ \est : \est = b (1+\epsilon)^i,~m \leq \est \leq 2km,~i \in \mathbb{Z} \} $,
  the \algoSieveStreaming retrurns the ($\frac{1}{2}-\epsilon$) approximation of
  the optimal result for $\query$.
\end{theorem}

In this way, we can shift the original estimations with a coeffecient and just
make sure the set of estimations covers the whole possible range of optimal
result. Therefore, instead of keep the $(1+\epsilon)^i$ estimations in advance,
we just shift all the estimations for a coeffecient of $b = e^(-2 \lambda \delta t)$
and generate some new estimations for optimal result of value $b(1+\epsilon)^i$.
In frequent update social graph, $b$ will decrease dramatically, and finally out
of the range and hard to detect.
In this we can adjust the base parameter to $b'$, which defined as
$b' = b  (1+\epsilon)^j$, where $j \in \mathbb{N}$ and is
chosen to make $b'$ most close to 1. The estimations therefore can be rewritten
as $b' (1+\epsilon)^{i-j}$.

To minimizing the times of calling of \algoTimeDecay, we use the exponential influence
instead of decaying. For example, we keep track of the base of time $t_0$, and use influence
based on $t_0$, i.e., the influence for action $\action = \langle u_r, t_r, v_e, t_e \rangle$ is
definded as $\infl{u_r \rightarrow u_e} = e^{\lambda  (t_r + t_e - 2  t_0)}$ instead
of $e^{\lambda (t_r + t_e - 2  t_{cur})}$, where $t_{cur}$ is the current timestamp.
When pushing result to user, we lazy decay the influence by $e^{-2\lambda (t_{cur} - t_0)}$.

The influence grows dramatically and is easily out-of-range. However,
we can predefine the threshold $\tau_{\syminfl}$. Just when influence grows larger than $\tau_{\syminfl}$ should
we call \algoTimeDecay to smaller down the influence data and set base of time $t_0 = t_{cur}$.

The overall process of \algoTimeDecay is shown in Algorithm~\ref{algo:timedecay}.

\input{algotimedecay}

\subsubsection{Lazy Time Decaying}

In frequent update social networks, there are thousands of timestamps, in a second,
which are usually represented as Unix timestamps.
In this case, we are not interested in the influence change before and after a
certain social action, e.g., a new post or a comment to a former post.
Therefore we will not call the subscription pushing process after every actions.

It's also unecessary for us to call \algoTimeDecay every time the timestamp change.
However, when pushing result to every subscriptions, we may temporarily decay to
fits the correct influence of based on $t_{cur}$.
The process will be much more time-consuming than calling \algoTimeDecay
when the number of subscriptions grows significantly.
So the best timing for us to call \algoTimeDecay process is just before subscription pushing.
In sum, we can define the time decay condition.

\begin{definition}
  \textbf{Time Decay Condition. } The timing to call \algoTimeDecay is before
  subscription pushing, or:
  \begin{equation}
    \max_{u \in \query} \infl{u} \geq \tau_{\syminfl}.
  \end{equation}
\end{definition}

%% file: sievestreaming.tex
\begin{figure}[!t]
\linesnumbered \SetVline\small
\begin{algorithm}[H]
\caption{\algoSieveStreaming} \label{algo:sievestreaming}
$E \leftarrow \{ e=(1 + \epsilon)^i:i \in \mathbb{Z}\}$ \;
For each $e \in E$, $S_e \leftarrow \emptyset$ \;
$m \leftarrow 0$ \;
\For{$i = 1 \mbox{ to } n$}
{
  $m \leftarrow \max(m, \infl{s})$ \;
  $E_i \leftarrow \{ e=(1 + \epsilon)^i:m \leq (1 + \epsilon)^i \leq 2km, i \in \mathbb{Z}\}$ \;
  Delete all $S_e$ such that $e \notin E_i$ \;
  \For{$e \in E_i$}
  {
    \If{$\margin{s}{S_e} \geq \frac{\frac{e}{2} - \infl{S_e}}{k - |S_e|}$}
    {
      $S_e \leftarrow S_e \cup \{ s\} $
    }
  }
}
\Return $\argmax_{e \in E_n} \infl{S_e}$ \;
\end{algorithm}
\vspace{-1.5em}
\end{figure}

%% file: algotimedecay.tex
\begin{figure}[!t]
\linesnumbered \SetVline\small
\begin{algorithm}[H]
\caption{\algoTimeDecay~($t_{cur}$)} \label{algo:timedecay}
// called when $\max_{u \in \query}\infl{u} \geq \tau_{\inf}$ \\
$d \leftarrow e^{-2 \lambda (t_{cur} - t_0)}$ \;
$b \leftarrow b d$ \;
$j \leftarrow$integer that let $b (1+\epsilon)^j$ closest to 1 \;
$b \leftarrow b (1+\epsilon)^j$ \;

\For{every $\est$ of every subscription $\query$}
{
  $m \leftarrow m  d$ \;
  Decay the $\est$ by $d$ \;
  Adjust estimations to cover the range of $(m, 2km)$ \;
}
\For{every estimation set $\estset$}
{
  Decay $\estset.inf$ by $d$ \;
}
\end{algorithm}
\vspace{-1.5em}
\end{figure}

%% file: prefixtree.tex
\section{Minimizing Times of Computation of Marginal Influence} \label{sec:prefixtree}

To support differnet query subscriptions, a naive solution is to maintain different
\algoSieveStreaming process for each sbuscription. As a result, the number of
estimations will grow linearly to the number of subscipritons, as well as the number
of candidate sets to calculate for each action update. However, some candidates
contains same users, and will cause a same candidate set to calculate multiple
times. For example, within an action a target user is $v$, for each same candidate
sets $S$, we should calculate the marginal influence for each candidate sets;
but as the marginal influence of a user $v$ w.r.t. to candidate set $S$ will not change,
we can just calculate the marginal influence only once. To accelerate action
update, we hope there is a data structure that maintains these same candidate sets
only once for each, in order to minimize the times of marginal influence computation.


Candidate sets also have \emph{downward closure property}, which means that some
of the conditions which a subset satisfies can be also applied to the any of its
supersets. Some of the properties are usefule in pruning out some of the unnecessary
marginal influence computation.
This properties are discussed in the following subsection~
\ref{subsec:purningconditions}.

When scanning through the action stream, new candidate sets will be generated,
while some of the exsisting candidate sets will be erased. To erase a candidate
set, it is required to search for a certain set in a collection of sets effeciently.
This means our data structure should support efficient search for set for a large
collection of sets.

We therefore propose a \emph{Prefix Tree Structure} to manage every condidate sets.
In this structure, each candidate set will be stored only once. The subsumption
relationship of differnet candidates are also maintained, in order to support pruning
over the collection of candidate sets using downward closue property. The insertion
and the erasion of candidates on the Prefix Tree is also efficient, which allows
us to manage the collection of ests dynamically.

In the following of this section, we will first explain how the Prefix Tree can
help us manage candidate sets and solve the \psim problem. Then we will discuss
the downward closure properties and the pruning conditions of marginal influence
computation, which can help us further minimize the times to calculate marginal
influence. In the end of this section, we will discuss how to maintain the structure
dynamically.

\subsection{Prefix Tree Structure and Action Update Process}

\begin{figure}[t!]
  \includegraphics[width=\linewidth]{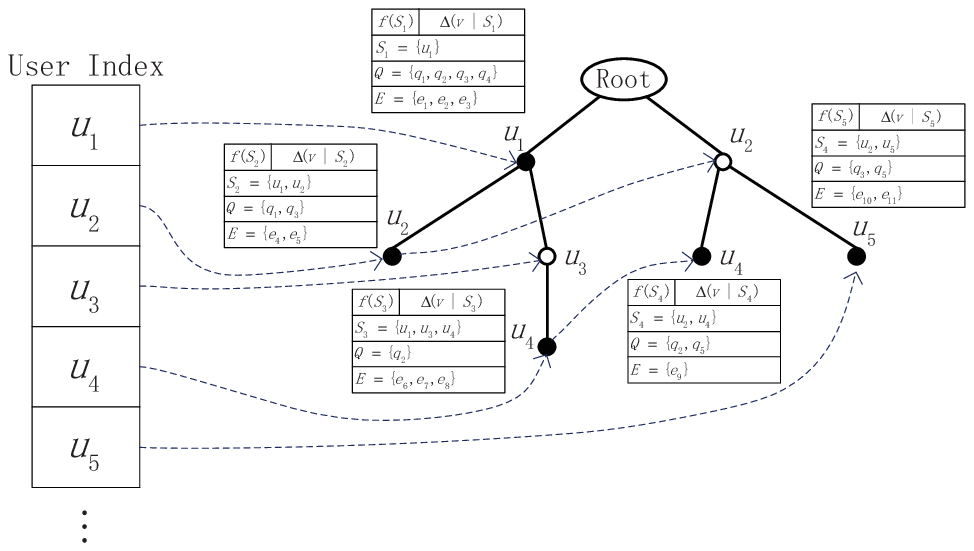}
  \caption{Prefix Tree Structure}
  \label{fig:prefixtree}
\end{figure}

In this part, we will frist introduce the Prefix Tree Structure. Then we will discuss
how we can use this structure to help us solve \psim prblem.

The Prefix Tree Structure is shown in fig~\ref{fig:prefixtree}, it has 2 parts:
the first part is a User Index; the second part is the prefix tree constructed by
elements in candidate sets. The User Index is a list, and each item of it is the
user id and the pointer to the location in the prefix tree where the user first appears.
The the location it points to will further points to where the user appears next time.
The linked list will continue until it contains all locations of the user. Each node
on the Prefix Tree is a user id, shose locations satisfy the following condition:
The user id from root to leaves is in increasing order.
A path from the root to solid points represents an candidate set, a solid
point, therefore, coresponded to a candidate set. Thus, we also refer a candidate
set $S$ as a \emph{path} on the Prefix Tree. The hollow points is not responded to
any candidate sets, but is an element in the paths it belongs to.
Each solid point restore the essential information related to the coresponded
candidate set. It has 5 main field: $\infl{S}$ stores the influence of the path,
$\margin{v}{S}$ contains the calculated marginal influence of user $v$ w.r.t.
current candidate set, $S$ contains all element on the path, and $Q$ and $E$ is
the related set of subscriptions and estimations, respectively.

The Prefix Tree contains all the essential information we need to solve \psim problem.
Here we will discuss how the structure deal with an action update and push reuslts
to every subscriptions.


When an action $\action  = \langle u_e, t_e, u_r, t_r \rangle$ arrives, the following
steps will be taken in sequence:

\begin{itemize}
  \item \textbf{Influence update. }
  Calculate the influence of action $\action$ as $\infl{u_r \rightarrow u_e}$,
  the influence of edge $\edge = \langle u_r, u_e \rangle$ will be also updated in
  social graph. The following update is to update the influence of candidate sets
  which contains $u_r$.
  \item \textbf{Calculate marginal influence. }
  We will calculate the marginal influence of target user $u_r$ w.r.t. to the
  corelated candidate sets, i.e. paths on the Prefix Tree.
  These candidates are all related to the subscriptions
  to which target user $u_r$ belongs to, as target user $u_r$ will never be inserted
  to those candidates which has no shared subscriptions with it. As the corelated
  set of subscriptions of a path $S$ is the subset of the set of subscriptions of
  its prefix paths, and the root element (which is empty set), is related to every
  subscriptions. We can calculate the marginal influence of each paths in Depth
  First Search (DFS) style. Once a path shares no commen subscriptions with user
  $u_r$, the we can prun out the subtree of the current path.
  As a result, the marginal influence of $u_r$ w.r.t. every related candidate sets
  will be calculated only once.
  \item \textbf{Judge sieve condition and update Prefix Tree. }
  After all the marginal influence is calculated, we further decide the candidate
  sets into which the target user can be inserted. For each candidate sets calculated
  in the previous step, we check the sieve condition for each of the coresponded
  estimations of the set. If there's a estimation $\est$ satisfies the sieve condition in
  candidate $S$, the new path $S' = S \cup \{ u_r\}$ will be created and inserted
  into the Prefix Tree, by which we will be further discussed in the following parts
  of this section. If the path $S'$ is already exist, the only thing we do it to
  move the estimation $\est$ from its current path to the new path; otherwise,
  the new path will be created in the Prefix Tree, and the same estimation movement
  will be performed. After a new path for a current path is generated, whether
  a existing path or a newly inserted path, the current path will point to the new path,
  in case of other following estimations can be moved quickly and without search
  for the path for multiple times. The information of the new path can be retrived
  from the current path, which will be further discussed in the following parts in
  this section. The erasion of the paths will be performed after
  the insertion. In this procedure, we check if there is still estimation linked to
  the candidate set. If no single estimation found, the path will be erased from
  the Prefix Tree. Instead of checking all paths on the Prefix Tree, we can only
  check those calculate in the previous step, as all the estimation removement is
  preformed within these candidates.
  \item \textbf{Push results to subscipritons. }
  In this step, we push the current influence of each path on the Prefix Tree
  to every subscriptions. For each path on the Prefix Tree, we push the influence
  of the path to all its related subscriptions. Within all the influence pushed
  to subscriptions, we pick out the largest influence among them as the result of
  the subscription. The result will be stored in the subscipriton for user to
  visit.
\end{itemize}

\begin{figure}[t!]
  \includegraphics[width=\linewidth]{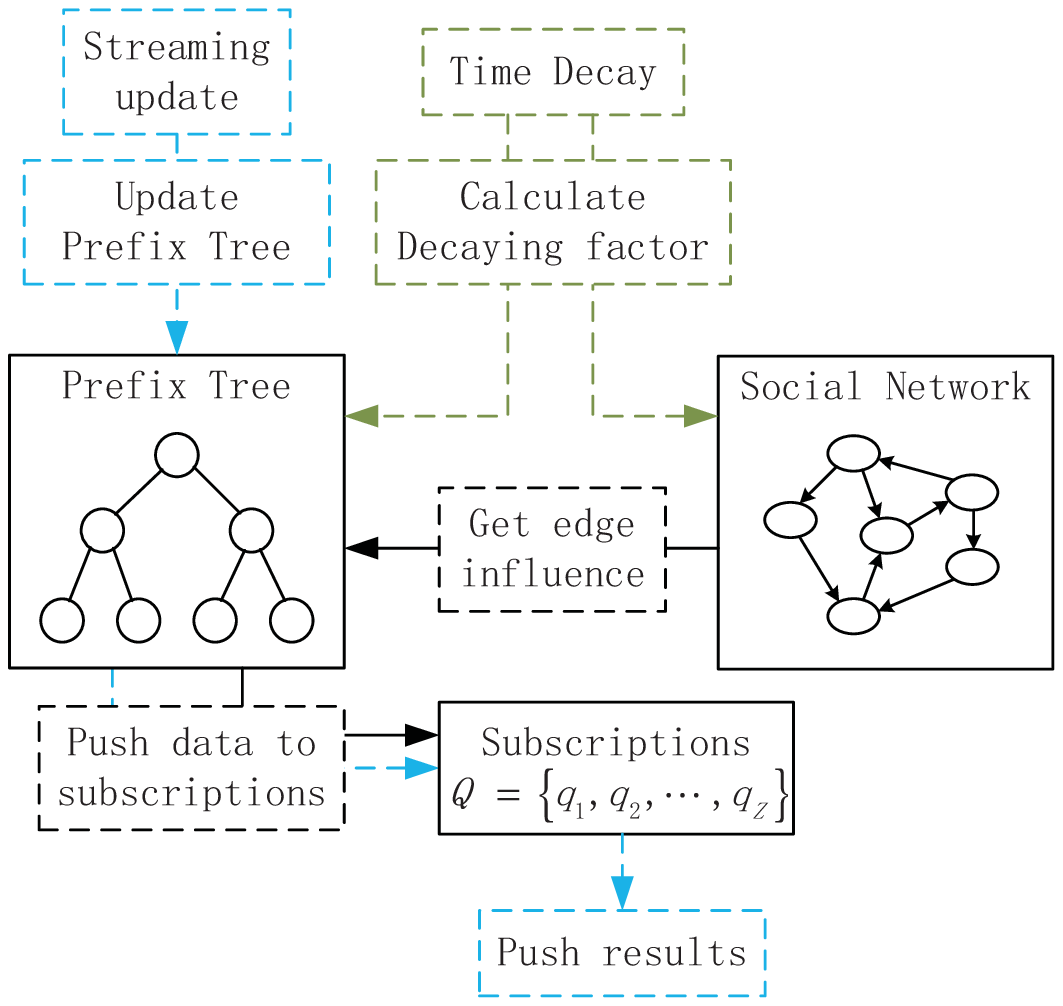}
  \caption{Updata and time decay process}
  \label{fig:flowchart}
\end{figure}

The overall procedure is summarized in Fig \ref{fig:flowchart}

We calculate all the marginal influence in one step and udpate the candidate sets
and Prefix Tree Structure in another step. And all the information is calculated
on the Prefix Tree, which is finally pushed to subscipritons.

In the following sections, we will further discuss how we continue minimizing the
times of marginal influence in step 2 and how we efficiently update the Prefix Tree
in step 3.

\subsection{Pruning Conditions}

Even we calculate the related candidates only once in every update, the times of
influence marginal calculation will grows dramatically when the number of subscriptions
grows. Moreover, the calculation method of marginal influence under different influence
model can be very different. We therefore aim at minimize the times we calculate
marginal influence. In this part, we will discuss 3 \emph{downward closure properties}
of Prefix tree, and propose three pruning conditions based on these properties to
prun out some branches of the Prefix Tree.

If an element $v$ is in one path $S$, for all paths $S'$ which satisfy
$S' \supset S$, it is clear that $v$ is also contained in $S'$.
On the other hand, for each node on the Prefix Tree $u$, its related path is contained
in the related paths of nodes of its subtree, called \emph{super-path} of path $S$.
Therefore, we can propose the first downward closure property:

\begin{definition}
  If one path of the Prefix Tree $S$ contains element $u$, all the super-path of
  this path also contains element $u$.
\end{definition}

With this property, we can prun out the subtree rooted to the target user $u_r$, as
the the paths related to the nodes on $u_r$'s subtree contains $u_r$, and therefore
don't need to decide whether the user can be inserted into the path. We therefore
reach our first pruning condition:

\begin{definition}
  \textbf{First Pruning Condition. }
  In the DFS process marginal influence
  computation for target influencer $u_r$ using DFS,
  we can prun down the subtree rooted to current tree node $u$ if $u = u_r$.
\end{definition}

This pruning conditian can help us prun out the nodes that are unnecessary to
calculate marginal influence. The second downward closure property which reveals
the relationship of paths can help us prun out the irrelated paths.

Noted that the related set of subscriptions of a give path is the intersection of
all the related subscriptions of all users in the path, that is,

\begin{equation}
  Q_S = \cap_{u \in S} Q_u
\end{equation}

where $Q_s$ and $Q_u$ are the set of related subscription of path $S$ and $u$,
respectively. The intuition is that, as a subscription is equivilent to a subset
of user set of a social network. Therefore, if a set $S$ belongs to subscription
$q$, it means that all the elements in set $S$ will be also included in the $q$,
otherwise the set can't be included in $q$. As a result, we can easily find out
the second downward closure property:

\begin{definition}
  Given path $S$, for each its super-path $S'$, we have $Q_S \supset Q_{S'}$.
\end{definition}

According to the relationship between a candidate set and its responded subscriptions,
we can know that, if the target user $u_r$ and a set $S$ shares no common subscription,
i.e., $Q_{u_r} \cap Q_{S} = \emptyset$, the set $S' = S \cup \{ u_r \} $ will not
belongs to any subscriptions. Therefore, we have our second pruning condition:

\begin{definition}
  \textbf{Second Pruning Condition. }
  If path $S$ and target user $u_r$ have no common subscipriton, i.e., $Q_S \cap Q_{u_r}
  = \emptyset$, super-paths of $S$ can be pruned.
\end{definition}

In this pruning process, to overcome the overhead computing large subscription
sets intersection, we keep track of the related subscriptions of target user $u_r$.
After calculating the intersection of the $Q'_{u_r} = Q_{u_r} \cap Q_S$, we
keep track of $Q'_{u_r}$ for further subscription sets computation.

If the root of each subtree $u$ knows the minimal related estimation of the subtree,
denoted as $e_{min}^{u}$, we have the following downward closure property:

\begin{definition}
  For two nodes $u$ and $v$, and $v$ corelated to a super-path of $u$'s corelated path,
  then $\est_{min}^{u} \leq \est_{min}^{v}$.
\end{definition}

According to \algoSieveStreaming, every node $u$ which can be inserted into the a
candidate set $S$ should be satisfy the following condition:

\begin{equation}
  \infl{u} \geq \frac{\frac{\est}{2} - \infl{S}}{k - |S|}.
\end{equation}

Therefore, if for the minimal estimation of u $\est_{min}$, we have $\infl{u} <
\frac{\frac{\est_{min}}{2} - \infl{S}}{k - |S|}$, we can skip the computation of
the marginal influence of this cadidate set.
Assuming that the subtree node knows the minimal estimation among the its subtree,
we have:

\begin{definition}
  \textbf{Third Pruning Condition. }
  For Prefix Tree node $u$ and the minimal estimation among the subtree rooted to u
  $\est_{min}^{u}$. We can prun out the subtree rooted to $u$ if
  \begin{equation}
    \margin{u}{S'} < \frac{\frac{\est_{min}^{u}}{2} - \infl{S}}{k - |S|}
  \end{equation}
  where $S'$ is the nearest sub-path of related path of $u$.
\end{definition}

\subsection{Prefix Tree Maintainance}

The insertion of user into candidate sets and the erasion of candidate sets will
cause the insertion of a new path and the delete of a existing path. In this part,
we will discuss how we update the Prefix Tree according to the marginal influence.

Path insertion happens when a new vertex satisfies the sieve
conditon and inserted into the candidate set. In this case, we should search through
the Prefix Tree to find if the path already exists. If such path has been found, the
only thing to do is to relink the estimation to a new path that related to it; otherwise
a new path should be inserted into the right place of the Prefix Tree, while the
relink should be done following the insertion. Algorithm~\ref{algo:modify} shows
the process of finding path and relink the estimation $\est$ and its new path.
When a new path is generated, the influence of the path will be retrived by adding
the marginal influence calculate in previous step to the influence of the original
candidate set. The related subscipritons of the new path is the intersection of the
subscipritons of original path and the related subscriptions of target user $u_r$.

In the \algoFindPath process, the algorithms searches the path related to the
candidate set $\estset$ and returns the path. The algorithm begins from the root node,
where it finds if there is a child $c$ of current node which is equal to the first element of
$\estset$. If found, it iteratively search the rest of the $\estset$ begins from
$c$, until if reaches the end of $\estset$. Otherwise, the algorithm inserts the
rest of $\estset$ as a branch from current node.

The clearance of a path happens when an candidate set has no related estimations.
The paths which should probably be erased is restrained to those the marginal influence
of which has been computed.
Candidate sets which satisfies the sieve condition will generate a new path in
the process of Algorithm~\ref{algo:modify} and move the related estimations to their
newly related path. Therefore, the links between estimations and candidate sets
changes dynamically. When the links to a candidate set, i.e., a path on the Prefix
Tree, it should be erased out from the Tree. The algorithm of clearing a estimation
set is shown in Algorithm~\ref{algo:clear}. The algorithm first check whether the
path is the longest path in its branch, i.e., the last node of the path is leaf. If
so, it will continue to earse the path starting from the leaf node, until it finds
first node which has more than one child or has related estimations to it; otherwise,
the the algorithm doesn't do anything, as the path is related to other candidate sets.

\input{modify}
\input{findpath}
\input{clear}

%% file: modify.tex
\begin{figure}[!t]
\linesnumbered \SetVline\small
\begin{algorithm}[H]
\caption{\algoModify~($u, \estset, \est$)} \label{algo:modify}
$\estset' \leftarrow \estset \cap \{ u \}$ ; // $\estset'$ is in sorted order \\
\If{new path haven't been generated}
{
  $S' \leftarrow$\algoFindPath~($root, \estset'$) \;
}
\Else{$S' \leftarrow the newly generated path$ \;}
$\infl{S'} = \infl{S} + \margin{u_r}{S}$ \;
$Q_{S'} \leftarrow Q_{S} \cap Q_{u_r}$ \;
Remove the link of $\est$ from $\estset$ and relink it to $S'$ \;
\end{algorithm}
\vspace{-1.5em}
\end{figure}

%% file: findpath.tex
\begin{figure}[!t]
\linesnumbered \SetVline\small
\begin{algorithm}[H]
\caption{\algoFindPath~($node, \estset$)} \label{algo:findpath}
$u \leftarrow $the first element of $\estset$ \;
$\estset \leftarrow \estset \setminus \{ u \} $ \;
\If{one of the child $c$ of $node$ s.t. $c=u$}
{
  \lIf{$\estset = \emptyset$}{\Return path from $root$ to $c$ \;}
  \Return \algoFindPath~($c, \estset$) \;
}
\Else{
  Insert $\estset$ sequentially as a branch of $node$ \;
  Insert new position of each newly insert node into vertex index \;
  \Return the new path \;
}
\end{algorithm}
\vspace{-1.5em}
\end{figure}

%% file: clear.tex
\begin{figure}[!t]
\linesnumbered \SetVline\small
\begin{algorithm}[H]
\caption{\algoClear~($\tpath$)} \label{algo:clear}
\For{each path whose marginal influence calculated}
{
  \lIf{the last element of $\tpath$ isn't leaf}{\Return \;}
  \While{the last element has no related estimations or has less than 2 children}
  {
    Erase the last element of $\tpath$ from Prefix Tree \;
    Erase the related position linkage in the vertex index \;
  }
}
\end{algorithm}
\vspace{-1.5em}
\end{figure}

%% file: experiment.tex
\section{Experiment} \label{sec:experiment}

\subsection{Experimental Setup}

\noindent \textbf{Datasets. }
We collect data of research papers from \emph{DBLP}, real world social network
\emph{Reddit} and \emph{Twitter}. The timestamp of DBLP dataset is based on the
publish year of research paper, it thus updates very slow, each timestamp contains
millions of actions. While Reddit and Twitter is frequent update social networks.
There are only tens of actions per timestamp.

\begin{itemize}
	\item \textbf{DBLP: } DBLP is a datasets of research papers retrived from DBLP
	and other sources~\cite{dblpdataset}. For each paper item, the dataset contains
	7 fields of information of the paper: \emph{paper id}, \emph{title}, \emph{authors},
	\emph{venue}, \emph{year}, \emph{references} and \emph{abstract}. The dataset
	contains 2,503,993 valid research papers from 1,422,578 authors. Keyword are extracted
	from the the title, venue and abstract fields of dataset based on TF-IDF.
	\item \textbf{Reddit: } Reddit is an online social forum with large number of
	users. The dataset is devided into \emph{posts} and \emph{comments}. The data collects
	the essential information about the response relationship and topics from which
	we can generate subscriptions. The dataset contains 1,995,836 users, 30,744,232 streaming
	actions and 2,554,003 subreddits, which is regarded as topic keywords.
	\item \textbf{Twitter: }
\end{itemize}

\begin{table}[h]
	\caption{Statistics on datasets}
	\label{table:datasets}
	\newcommand{\tabincell}[2]{\begin{tabular}{@{}#1@{}}#2\end{tabular}}
	\resizebox{\linewidth}{!}{
	\begin{tabular}{|c|c|c|c|c|c|}
		\hline
		Dataset & Vertex & Edges & Avg. Deg. & \tabincell{c}{Avg. Keywords \\ per User} & Stream \\ \hline
		DBLP    & 1,422,578 & 131,878,718 & 92.70 & 9.49 & 240,940,225 \\ \hline
		Reddit  & 1,995,836 & 24,537,116  & 12.29 & 3.82 & 30,744,232  \\ \hline
		Twitter & 0         & 0           & 0 & 0     & 0      \\
		\hline
	\end{tabular}
	}
\end{table}

The statistics of the abovementioned dataset are shown in Table~\ref{table:datasets}.

The subscriptions are generated by randomly choose a sample from keywords set of
the social networks. In real world subscription queries, users are usually interested
in a small area of topics compared to the total topic sets of the social networks.
On the other hand, using TF-IDF, we generate user subscriptions based on the informaton
provided by each keyword. We generate subscription sets that each user will be
related to 1.?? subscriptions in average.

\noindent \textbf{Approaches. }

\begin{itemize}
	\item \textbf{\IC: } \IC is a state-of-the-art dynamic IM algorithm. We assign
	an index to each of the actions sequentially in the social stream, and reconstruct
	according to the trigger model which is used by \IC.
	\item \textbf{\SIC: } \SIC is also proposed by \cite{SIM}. The parameter $\beta$,
	which controls the trade-off between quality and efficieny, is set to the proposed
	default value, i.e., $\beta = 0.2$.
	\item \textbf{\algoPSIM: } The \algoPSIM proposed in section \ref{sec:prefixtree}.
	We set the approximating parameter in \algoSieveStreaming $\epsilon$ to 0.1.
\end{itemize}

(How to compare multi-topic settings and non-topic setting???)

\noindent \textbf{Quality Matric. }
\IC and \SIC retrive the most influential users based on cardinality functions,
while our approach is based on one-hop coverage influence model described in
section~\ref{sec:problem}. In order to verify the quality of our solution, we
adopt well-reganized \emph{IC} influence model in evaluation the influence spread
of the results retrived by each approcah. For sliding window based approaches, we
use the default window size of 100K actions length. The solutions are retrived by
every timestamp and evaluated by IC influence model. Finally, we use the average
influence spread of all results as the quality metric.

\noindent \textbf{Performance Matric. }
We use \emph{throughput} as our performance matric. The throughtput is measured by
measuring the CPU time elapse of returning $L$ results, and divide $L$ by the
elapse time.

\noindent \textbf{Parameters. }
The parameters examined in our experiments:
(1) $k$ is the size of seed set.
(2) $\lambda$ is the time decaying constant controlling the weights decay of actions
over time elapse.
(3) $|Q|$ is the number of subscription queries to answer.
The summary of the parameters is listed in Table~\ref{table:params} with default
values in bold.

\begin{table}[h]
	\caption{Parameters in experiments}
	\label{table:params}
	\begin{tabular}{|c|c|}
		\hline
		Parameter & Values \\ \hline
		$k$       & 5, 25, \textbf{50}, 75, 100 \\ \hline
		$\lambda$ & 0.01, 0.05, \textbf{0.1}, 0.3, 0.5 \\ \hline
		$|Q|$     & \textbf{200K}, 400K, 600K, 800K, 1M \\
		\hline
	\end{tabular}
\end{table}

\noindent \textbf{Experemnt Environment. }
All experiments are conducted on a server machine running Ubuntu 14.04 with (CPU)
and 250 BG memory. \IC and \SIC is implemented in Java 8, while the implementation
of \algoPSIM is in C++.

\subsection{Performance of Time-Decaying \algoSieveStreaming}

We first test the efficiency and the performance of the time-decaying model, comparing
to the results of sliding window. In this part, we run naive \algoSieveStreaming
over a window size of 500K and \algoPSIM over DBLP dataset. In the case study of
\texttt{Neural Networks}, we search for the $k$-influencers
in the area of \texttt{Neural Networks}, and compare the influencers extracted by
the two influence model. We also compare the influence spread of users extracted
by the two models. Then we test the effeciency of the naive time-decaying
\algoSieveStreaming and sparse and lazy time-decaying \algoSieveStreaming when
varying the time decaying constant $\lambda$.

\subsubsection{Result Comparision between Time-Decaying and Sliding Window Models}

In this part, we will compare the performance of the time-decaying model and sliding
window models.

\begin{table}
	\caption{Well-known influencer of \texttt{Neural Networks}}
	\label{table:casestudy}
	\begin{tabular}{|c|c|c|}
		\hline
		Year & Time-decaying & Sliding window \\ \hline
		2000 & Michael I. Jordan & \ldots \\ \hline
		2010 & \ldots & \ldots \\
		\hline
	\end{tabular}
\end{table}

\noindent \textbf{Case Study on \texttt{Neural Networks}: }
The result of the case study of extracting $k$-influencers under the topic of
\texttt{Neural Networks} is shown in table \ref{table:casestudy}.
It is shown that the time-decaying model can extract more well-known influencers
of the area of \texttt{Neural Networks} than sliding window do.
It is caused by the fact that sliding window throws the information of previous
actions. At the time of returning result for the most influencers, just few actions
performed by influencers are in the window. On the other hand, as there are increasing
number of research papers published each year, the sliding window of of fixed length
in different year includes actions diversed length of time, it therefore cannot
reveals the ture influencers over time.

\noindent \textbf{Quality}
(Influence spread of users extracted by the two models)

\subsubsection{Varying Decaying Constant $\lambda$}

In this part, we compare the effeciency of the naive \algoSieveStreaming and
sparse and lazy time-decaying \algoSieveStreaming.

\noindent \textbf{Throughput}

\subsection{Comparison of Differnet Approaches}

\subsubsection{Testing Influence Spread over $k$}

\subsubsection{Testing Throughput over $k$}

\subsection{Comparing with Other Approaches}

Campare the average update time per subscription of \algoPSIM and other algorithmes.

\subsection{Scalability}

Test throughput on number of subscriptions.